\documentclass{article}

\PassOptionsToPackage{numbers, compress}{natbib}

\usepackage[preprint]{neurips_2023}

\usepackage[utf8]{inputenc} 
\usepackage[T1]{fontenc}    
\usepackage{hyperref}       
\usepackage{url}            
\usepackage{booktabs}       
\usepackage{amsfonts}       
\usepackage{nicefrac}       
\usepackage{microtype}      
\usepackage{xcolor}         
\usepackage{graphicx}
\usepackage{textcomp}
\usepackage{subcaption}

\usepackage{gensymb}

{}
{}

\title{AI4GCC - Team: Below Sea Level \\Critiques and Improvements}

\author{%
Bram Renting*
\\
Leiden University \\ Delft University of Technology
\And 
Phillip Wozny*
\\
Tilburg University \\ Vrije Universiteit Amsterdam 
\AND 
Robert Loftin \\
Delft University of Technology
\And 
Claudia Wieners\\
Utrecht University
\And
Erman Acar \\
University of Amsterdam 
}

\begin{document}

\maketitle

\begin{abstract}
We present a critical analysis of the simulation framework RICE-N, an integrated assessment model (IAM) for evaluating the impacts of climate change on the economy. We identify key issues with RICE-N, including action masking and irrelevant actions, and suggest improvements such as utilizing tariff revenue and penalizing overproduction. We also critically engage with features of IAMs in general,  namely overly optimistic damage functions and unrealistic abatement cost functions. Our findings contribute to the ongoing efforts to further develop the RICE-N framework in an effort to improve the simulation, making it more useful as an inspiration for policymakers. 

\end{abstract}

\section{Introduction}

Over the course of implementing the first two submissions, we engaged with the simulation framework RICE-N. We interrogate both RICE-N as implemented by \citet{zhang2022ai} and the assumptions inherent to the class of models to which it belongs, known as integrated assessment models (IAMs). 

\paragraph{Executive summary}

Our key points of criticism can be categorically summarised as follows:

\begin{itemize}
    \item Action masking inflates model performance
    \item Issues with RICE-N:
    \begin{itemize}
        \item Most actions are irrelevant for key climate and economic indices.
        \item Tariffs do not impact reward of the affected state.
        \item Trade does not impact reward of the affected state as intended.
        \item Damages have little impact on reward.
    \end{itemize}
    \item Suggestions for RICE-N improvement:
    \begin{itemize}
        \item Use tariff revenue.
        \item Penalize overproduction.
        \item Allow for technology sharing and wealth redistribution.
        \item Let abatement cost depend on the previous mitigation level.
        \item Strengthen the damage function.
    \end{itemize}
    \item Issues with IAMs:
    \begin{itemize}
        \item The damage function is overly optimistic.
        \item Abatement costs do not depend on previous mitigation.
    \end{itemize}
\end{itemize}


\section{Problems with Action Masking}\label{sec:masking}

Action masking is used to disable actions that are inconsistent with negotiated agreements. Excessive masking can force protocols into seemingly desirable behavior despite being the result of random behavior. This phenomenon is evident in the Bilateral Negotiator released with the competition. The protocol performs better after training for one episode than after extensive training. This is due to the following well-intended design features. First, agents commit to the maximum accepted proposal. With 27 agents sending and receiving proposals, one of the 54 total proposals likely corresponds to a high mitigation rate. Second, masks are used to enforce the maximum accepted proposal. As such, high levels of mitigation are almost certainly enforced.

Finally, we argue that actions rendered inaccessible through masking are unrealistic as states can never be forced to commit to their agreements. Instead, collectively unfavorable actions should be made less desirable through extrinsic factors imposed by other states.


\section{Problems with the RICE-N model}

\subsection{Influence of actions on key measures}\label{sec:interconnectedness}
\begin{figure}
    \centering
    \includegraphics{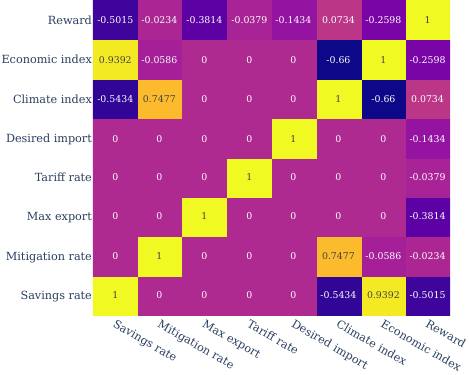}
    \caption{Correlation between RICE-N actions and performance measures. As can be seen, climate and economic indexes only correlate with mitigation and savings rates. Reward is more connected to actions.}
    \label{fig:correlation}
\end{figure}

In order to explore the space of attainable outcomes in RICE-N we now analyze the correlation of key performance measures to a range of possible action inputs. In this context, we restrict the actions such that they are fixed for a full simulation and that all regions perform the same action. We consider RICE-N with 10 discrete action levels. There are 5 different types of actions of which we want to sample every possible combination, resulting in a total of $10^5$ environment rollouts. We collected the climate index, economic index, and reward of the episode and generated the correlation matrix shown in \autoref{fig:correlation}.

\begin{figure}
    \centering
    \includegraphics{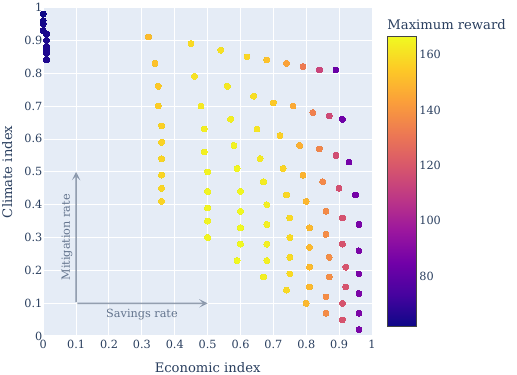}
    \caption{Results of manually sampling the RICE-N model with fixed actions. All regions perform identical actions for all timesteps. Every possible combination of actions is attempted and represented by a dot in this figure. The dots perfectly overlap in 100 points changing position only based on mitigation and savings rate. Of all the overlapping dots, the maximum reward of that point is visualized by a color gradient.}
    \label{fig:climate_vs_economic}
\end{figure}

We observe that there is practically no correlation between desired import, tariff rate, maximum export, and climate/economic index. Only the mitigation rate and savings rate have an impact on the economic index and the climate index. Realize that this makes all the other actions irrelevant for the first submission track of this competition. We further explore this by plotting the results of every episode in \autoref{fig:climate_vs_economic} and observe that the $10^5$ episodes form $100$ perfectly overlapping dots. The difference between dots along the $x$-axis represents the savings rate and along the $y$-axis the mitigation rate. We also observe that optimizing for reward without any restriction results in a low score both in the economic and climate index. Finally, we would like to note that gross output, and thus the economic index, is an internal affair that the states will not punish each other over. It is unrealistic to expect states to force themselves to increase their gross output as this would only lower their reward. Such issues make high-economic index solutions not learnable.

This is not the case for climate mitigation as states do have the incentive to punish other states that do not mitigate. There are two methods to punish non-compliant states: (1) imposing tariffs on those states, and (2) limiting imports from that state. However, \autoref{fig:correlation} suggests that there is barely an effect of tariffs on the reward. Even more surprising is the negative correlation between desired imports and the reward, which suggests that limiting imports from a state is rewarding that state. We study this in more depth in the following sections.

\subsubsection{Effects of tariffs on reward}\label{sec:tariffeffects}
The first sanctioning mechanism of the RICE-N model  is the tariff. This is informed by literature on climate economics, which uses tariffs and levies to adjust the cost of carbon~\cite{overland2022climate, nordhaus2015climate}. In previous IAMs, there is a static parameter corresponding to the loss of welfare per unit of tariff~\cite{nordhaus2021dynamic}. At each step, agents compare the cost of mitigation to the expected loss of welfare due to tariffs. Once the latter exceeds the former, mitigation becomes the preferred action.

Currently, there is no global authority that can externally sanction defection. As such, it is the responsibility of states to sanction one another. The organizers' implementation of the RICE-N model acknowledges this fact and grounds the sanction in the trade dynamics of the simulation itself~\cite{zhang2022ai}. As a result of that, there is no ``loss of welfare due to tariff'' parameter. It is assumed that if agents apply tariffs to each other it will result in a loss of revenue. However, as we will show empirically in the following analysis, tariffs have negligible to no impact on reward in most cases.

\begin{figure}
    \centering
    \includegraphics[scale=.6]{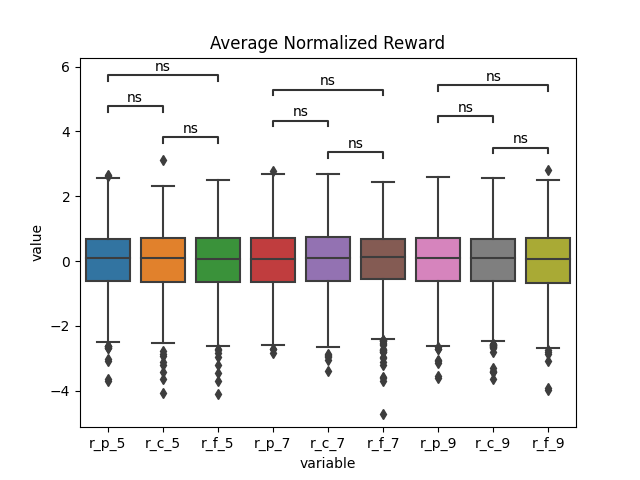}
    \caption{Comparison of the average normalized reward in each experimental condition: pariah, control, and free trade. The pariah agent receives a fixed tariff, the control agent follows the protocol, and the free trade agent has no tariffs. We examine three different fixed tariff values, 5, 7, and 9. The black diamonds represent outliers.}
    \label{fig:anr}
\end{figure}

\begin{figure}
    \centering
    \includegraphics[scale=.6]{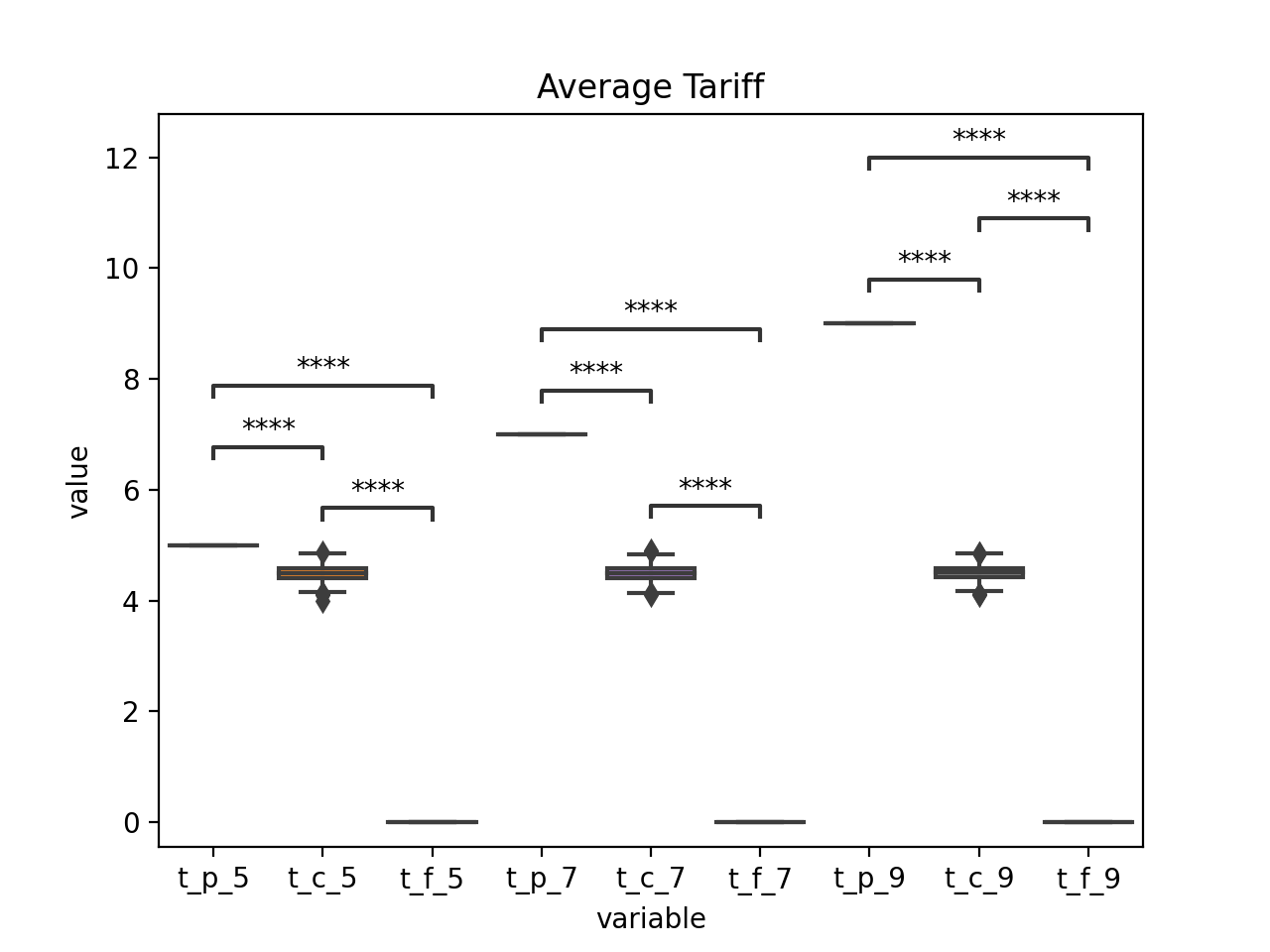}
    \caption{
    Visualization of the average tariff rate of the agent in question confirms that our experimental conditions work as intended.}
    \label{fig:ats3}
\end{figure}

\paragraph{Methods}

To measure the effects of tariffs on reward and trade dynamics, we conducted the following experiment comparing three agent groups: pariahs, controls, and free trade. Pariah agents received a fixed tariff from all other agents, control agents simply followed their trained protocol, and free trade agents received no tariff. We examined three different fixed tariff values for the pariah agents; 5, 7, and 9. Using a model trained without negotiation for $100\,000$ episodes, we ran the simulation 1000 times for each group and each experimental condition. 

During each iteration of the experiment, an agent was chosen uniformly at random as the subject. Between each group and each fixed tariff value, the agent and environment were reset. As such, each experimental condition contains data from all agents. As the reward varies considerably between states and is relatively stable within states, we normalize the reward by each state id. Therefore, we compare the effect of tariffs on rewards relative to each state.

In a follow-up experiment, we sampled the RICE-N model with fixed actions that represent an ideal trade scenario where the mitigation rate is $0.9$, savings rate $0.3$, desired import $0.9$, and maximum export $0.9$. Under such high trade conditions, the effect of tariffs would be expected to be magnified. We compared the rewards of each state with both maximum tariffs and no tariffs at all in order to measure the effect of tariffs on reward.

\paragraph{Results}
As visible in \autoref{fig:anr}, manipulating tariff magnitude does not result in significant differences in relative reward. To ensure that our tariff manipulation is effective, we also gathered the average tariff per subject in each treatment condition (see \autoref{fig:ats3}). Taken together, it indicates neither the maximum tariff nor the absence of tariff from all agents towards any single agent impacts that agent's reward. This suggests a flaw in the trade component of the reward calculation.

The second experiment showed that the reward of a state is untouched by other states' tariff actions. Even more surprising is that the reward of the state that imposed the tariff decreases by $0.02-6.50\%$.

\paragraph{Explanation}
For the RICE-N model to accurately model interstate commerce and climate-related tariff mechanisms, the tariff mechanism  needs to impact reward. The failure of tariffs to impact the reward can be traced back to how trade influences the reward which we explain in the following.

The reward is calculated as aggregate consumption which can be decomposed into two additive terms, foreign and domestic consumption, and trade is the amount of import/export from one country to another. Critically, trade takes two forms, scaled imports, and tariffed imports. The former is the overlap between the desired imports of one state corrected for gross output and the exporting capacity of the other. The latter adjusts the scaled imports by the inverse of the tariff. Foreign consumption utilizes tariffed imports and domestic consumption utilizes scaled imports. Therefore, if state A tariffs state B, it is only factored into the reduction of state A's foreign consumption term. State B's domestic consumption, which depends on its export to State A, remains unchanged by the tariff. Therefore, tariffing a state does not affect the reward of the state being tariffed.

\subsubsection{Effects of desired imports on reward}
We now focus on the desired imports and maximum exports actions, or simply the amount of trade. We sampled the RICE-N model with fixed actions where the mitigation rate is $0.9$, savings rate $0.3$, and tariffs $0$ to maximize the benefit of trade. We varied the desired imports and maximum export actions, observing the reward.

Reward increases for all states when trade is limited. If we compare the extreme cases of maximum trade and no trade, the reward per state increases by a percentage of $1-2322\%$, depending on the state, when no trade is happening. It appears that limiting imports cannot be used to negatively impact the reward of another state, but actually positively impacts it. This is likely the case because domestic consumption is preferred over foreign consumption in the current setup of RICE-N. As export lowers your domestic consumption and the import does not compensate for this, states will prefer not to export.

\subsubsection{State punishment potential}
We can conclude that neither imposing tariffs nor limiting imports are suitable sanction mechanisms in RICE-N. Moreover, limiting imports even increases the reward of the subjected state. As such, states have no leverage with which to negotiate. Any observed form of ``learned'' negotiation is likely a result of random behavior or favorable masking or both. Any optimal learning agent will end up at the point of maximum reward (see \autoref{fig:climate_vs_economic}) at a climate index of 0.33 and an economic index of 0.6. Therefore, the only mechanism to obtain more optimal policies is to exploit the RICE-N framework itself.

\subsection{Our submission one}
\begin{figure}
    \centering
    \includegraphics[scale=.5]{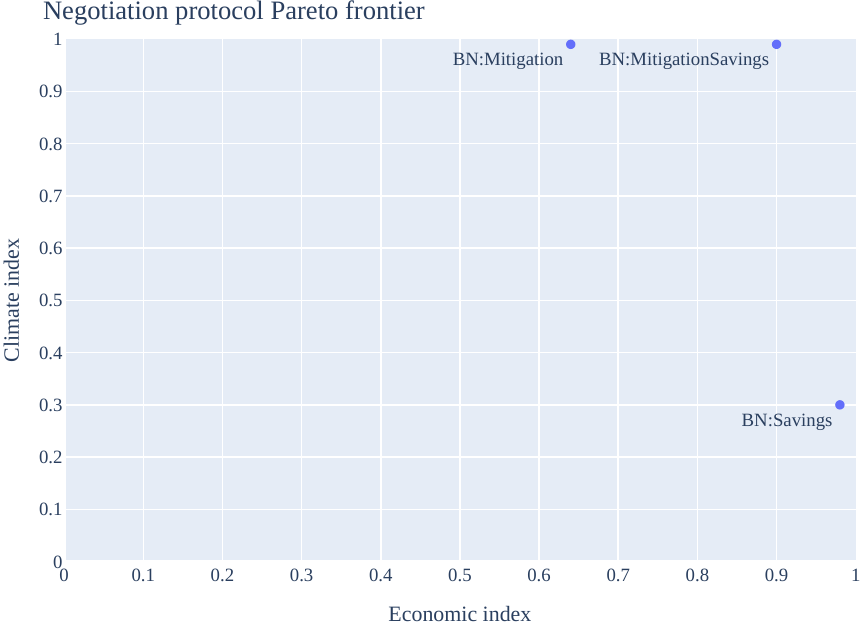}
    \caption{We compare three variations on the Bilateral Negotiator protocol. The name of each protocol indicates the action which is bilaterally negotiated. Each of these protocols was trained for one episode. Note, these three submissions are the same used for Track 1.}
    \label{fig:gt}
\end{figure}

Due to these shortcomings, we \emph{strategically} utilized three Bilateral Negotiator variations which make use of the action masking quirk described in \autoref{sec:masking}. That is, the protocols were essentially untrained. One protocol made use of favorable masking for savings, another for mitigation, and the third for both savings and mitigation. The resulting Pareto frontier is visible in \autoref{fig:gt}. We are aware of the fact that such protocols are not in the spirit of the competition. However, it was stated during the first workshop session that tracks one and two submissions may be independent and that the goal of track one is purely metric optimization.

\subsection{Finite horizon}
When no mitigation is performed by states, the damages reduce the gross output of states by $\sim8.5\%$. This is only a minor deduction of the gross output and does not reflect the real-world \emph{sword of Damocles} which is a climate disaster. From the perspective of the agents, the game simply ends and no large damages are achieved. Reducing an infinite horizon game to a finite one changes optimal strategies from a game-theoretical perspective.

We extended the horizon of the simulation to 200 and 300 years to let damages influence the agents more heavily. We trained agents without a protocol with the hope that major future damages would cause the agents to learn to mitigate. However, while damages rose to $\sim 13\%$ and $\sim 22\%$, mitigation rates remained as before. We also saw earlier in \autoref{fig:correlation} and \autoref{fig:climate_vs_economic} that reward negatively correlates with gross output (economic index). We can conclude that there is little effect of temperature rise on state reward.



\section{Suggestions for improvement}

\subsection{Tariff revenue}

Currently, tariff revenue is calculated during the climate economic simulation step but not used. Tariff revenue could contribute to government balance, the debt ratio, and the subsequent import capacity for the next climate economic simulation step. Literature on climate policy-induced economic inequality suggests that carbon tariff revenue should be reinvested in developing countries to develop climate mitigation infrastructure~\cite{goldthauHowOpenClimate2022}. 


\subsection{Cost of overproduction}

As stated in \autoref{sec:tariffeffects}, tariffs do not impact reward. A possible workaround could make use of scaled and tariffed imports. The differences between the two correspond to the amount of output produced for export but not ultimately consumed by the importer. Currently, the exporting country faces no penalty for overproduction. Adding the difference between tariffed imports and scaled imports to the export term of domestic consumption would lower overall consumption and function as a penalty to the country being tariffed. 


\subsection{Disaster influenced reward}
Future damages should have an impact on the reward of states, which seems to be limited currently. The presence of a horizon in the simulation creates a situation without a doomsday scenario. Simulating an infinite game is of course impossible, but adding an artificial \emph{sword of Damocles} might help. This could be in the form of a high negative reward when a certain temperature rise is passed.

\section{Issues with IAMs}

\subsection{Damage by climate change}

DICE's damage function describes the fraction of GDP lost to climate damage. It is unlikely that catastrophic climate conditions of a 5\degree C temperature rise would only result in a 5\% loss in GDP~\cite{woillez2020economic}. As such, more realistic damage functions are required, such as \citet{weitzman2012ghg} which allows for an ``infinitely bad'' climate, or \citet{burke2015global}, which attempts to construct a damage function from empirical data. 

Damage functions can also be viewed as secondary to the ``guardrail approach'', which emphasizes a target warming threshold. Policymakers then focus on staying under the threshold as opposed to avoiding damages~\cite{stern2022economics}


\subsection{The dynamics of mitigation costs}

Currently, mitigation costs persistently depend only on the current mitigation level. In reality, mitigation costs are transitional. Constructing a wind farm is costly during construction, but not so once completed. Persistent mitigation costs incentivize the latter investment, once abatement costs are decreased due to technology. In contrast, transitional mitigation costs incentivize early investment, as that reduces the cost of subsequent mitigation.

\citet{grubb2021modeling} proposed a correction to DICE which includes a transitional mitigation cost function. This allows states to negotiate around rates of change as opposed to absolute mitigation levels which is more in line with real-world climate targets (e.g.~halving emissions by 2030).

\begin{ack}
We would like to thank Maikel van der Knaap, Cale Davis, Albert Bomer, Catholijn Jonker, and Holger Hoos for their time spent discussing various topics of this competition.

This research was (partly) funded by the \href{https://hybrid-intelligence-centre.nl}{Hybrid Intelligence Center}, a 10-year programme funded by the Dutch Ministry of Education, Culture and Science through the Netherlands Organisation for Scientific Research, grant number 024.004.022.
\end{ack}

\bibliographystyle{IEEEtranN}
\bibliography{ref}

\begin{thebibliography}{10}
\providecommand{\natexlab}[1]{#1}
\providecommand{\url}[1]{#1}
\csname url@samestyle\endcsname
\providecommand{\newblock}{\relax}
\providecommand{\bibinfo}[2]{#2}
\providecommand{\BIBentrySTDinterwordspacing}{\spaceskip=0pt\relax}
\providecommand{\BIBentryALTinterwordstretchfactor}{4}
\providecommand{\BIBentryALTinterwordspacing}{\spaceskip=\fontdimen2\font plus
\BIBentryALTinterwordstretchfactor\fontdimen3\font minus
  \fontdimen4\font\relax}
\providecommand{\BIBforeignlanguage}[2]{{%
\expandafter\ifx\csname l@#1\endcsname\relax
\typeout{** WARNING: IEEEtranN.bst: No hyphenation pattern has been}%
\typeout{** loaded for the language `#1'. Using the pattern for}%
\typeout{** the default language instead.}%
\else
\language=\csname l@#1\endcsname
\fi
#2}}
\providecommand{\BIBdecl}{\relax}
\BIBdecl

\bibitem[Zhang et~al.(2022)Zhang, Williams, Phade, Srinivasa, Zhang, Gupta,
  Bengio, and Zheng]{zhang2022ai}
T.~Zhang, A.~Williams, S.~Phade, S.~Srinivasa, Y.~Zhang, P.~Gupta, Y.~Bengio,
  and S.~Zheng, ``Ai for global climate cooperation: Modeling global climate
  negotiations, agreements, and long-term cooperation in rice-n,'' \emph{arXiv
  preprint arXiv:2208.07004}, 2022.

\bibitem[Overland and Huda(2022)]{overland2022climate}
I.~Overland and M.~S. Huda, ``Climate clubs and carbon border adjustments: a
  review,'' \emph{Environmental Research Letters}, vol.~17, no.~9, p. 093005,
  2022.

\bibitem[Nordhaus(2015)]{nordhaus2015climate}
W.~Nordhaus, ``Climate clubs: Overcoming free-riding in international climate
  policy,'' \emph{American Economic Review}, vol. 105, no.~4, pp. 1339--70,
  2015.

\bibitem[Nordhaus(2021)]{nordhaus2021dynamic}
------, ``Dynamic climate clubs: On the effectiveness of incentives in global
  climate agreements,'' \emph{Proceedings of the National Academy of Sciences},
  vol. 118, no.~45, p. e2109988118, 2021.

\bibitem[Goldthau and Tagliapietra(2022)]{goldthauHowOpenClimate2022}
A.~Goldthau and S.~Tagliapietra, ``How an open climate club can generate carbon
  dividends for the poor,'' \emph{Bruegel-Blogs}, 2022.

\bibitem[Woillez et~al.(2020)Woillez, Giraud, and Godin]{woillez2020economic}
M.-N. Woillez, G.~Giraud, and A.~Godin, ``Economic impacts of a glacial period:
  a thought experiment to assess the disconnect between econometrics and
  climate sciences,'' \emph{Earth System Dynamics}, vol.~11, no.~4, pp.
  1073--1087, 2020.

\bibitem[Weitzman(2012)]{weitzman2012ghg}
M.~L. Weitzman, ``Ghg targets as insurance against catastrophic climate
  damages,'' \emph{Journal of Public Economic Theory}, vol.~14, no.~2, pp.
  221--244, 2012.

\bibitem[Burke et~al.(2015)Burke, Hsiang, and Miguel]{burke2015global}
M.~Burke, S.~M. Hsiang, and E.~Miguel, ``Global non-linear effect of
  temperature on economic production,'' \emph{Nature}, vol. 527, no. 7577, pp.
  235--239, 2015.

\bibitem[Stern et~al.(2022)Stern, Stiglitz, and Taylor]{stern2022economics}
N.~Stern, J.~Stiglitz, and C.~Taylor, ``The economics of immense risk, urgent
  action and radical change: towards new approaches to the economics of climate
  change,'' \emph{Journal of Economic Methodology}, vol.~29, no.~3, pp.
  181--216, 2022.

\bibitem[Grubb et~al.(2021)Grubb, Wieners, and Yang]{grubb2021modeling}
M.~Grubb, C.~Wieners, and P.~Yang, ``Modeling myths: On dice and dynamic
  realism in integrated assessment models of climate change mitigation,''
  \emph{Wiley Interdisciplinary Reviews: Climate Change}, vol.~12, no.~3, p.
  e698, 2021.

\end{thebibliography}

\end{document}


\begin{appendices}

\section{Training}
\label{appendix:training}

\begin{table}[H]
\centering
\caption{Training configuration}\label{tab:config}
\begin{tabular}{ll}
\toprule
Parameter                      & Value                                \\ \midrule
number episodes                & 100000                               \\
batch size in episodes         & 60                                   \\
framework                      & torch / rllib                        \\
value function loss coefficient & 0.1                                  \\
entropy coefficient schedule   & (0, 0.5), (40000, .1), (70000, 0.05) \\
clip grad norm                 & TRUE                                 \\
max grad norm                  & 0.5                                  \\
gamma                          & .92                                  \\
learning rate                  & 0.0005                               \\ \bottomrule
\end{tabular}
\end{table}

\section{Formalization}
\label{appendix:formalization}
For the negotiation protocols referenced in this submission, we utilized the configuration referenced in \autoref{tab:config}. Furthermore, we utilized a modified masking mechanism in the model to mask out all actions not relevant to a given step. We do this to stabilize training.

In addition to the formalism stated in the original white paper, given $n$ agents (i.e., $[n] = \{1, \ldots, n\}$)\footnote{In the simulations, $n = 27$}, action space $k$ has two subspace crucial to our  protocol 
\begin{itemize}
    \item  \textbf{proposal} $p_i \in \{0, \ldots, 9\}$ which corresponds to the  mitigaton level  that every agent $i \in [n]$ chooses for the climate club it takes place. 

    \item \textbf{evaluation} $\mathbf{e}_i =   \langle e_1, \ldots, e_{n} \rangle$   is a vector representing \emph{evaluation} of agent $i$ for all the other agents; that is,  each entry $e_j \in \{0, 1\}$ (with $j \in [n]$) denotes that the agent $i$ either accepts (i.e., $e_j = 1$) or rejects (i.e., $e_j = 0$) the proposal from agent $j$. \footnote{Here, to keep things simple, we ignored the case where $i=j$.}
\end{itemize}

\end{appendices}